\documentclass[sigconf]{acmart}
\AtBeginDocument{%
  }

\setcopyright{acmlicensed}
\copyrightyear{2018}
\acmYear{2018}
\acmDOI{XXXXXXX.XXXXXXX}

\acmConference[Conference acronym 'XX]{Make sure to enter the correct
  conference title from your rights confirmation email}{June 03--05,
  2018}{Woodstock, NY}
\acmISBN{978-1-4503-XXXX-X/18/06}

\copyrightyear{2024}
\acmYear{2024}
\setcopyright{acmlicensed}\acmConference[VINCI 2024]{The 17th International Symposium on Visual Information Communication and Interaction}{December 11--13, 2024}{Hsinchu, Taiwan}
\acmBooktitle{The 17th International Symposium on Visual Information Communication and Interaction (VINCI 2024), December 11--13, 2024, Hsinchu, Taiwan}
\acmDOI{10.1145/3678698.3687186}
\acmISBN{979-8-4007-0967-8/24/09}

\begin{document}


\title[Memory Remedy]{Memory Remedy: An AI-Enhanced Interactive Story Exploring Human-Robot Interaction and Companionship}

\author{Lei Han}
\authornote{Both authors contributed equally to this research.}
\email{lhan229@connect.hkust-gz.edu.cn}
\orcid{0009-0001-7157-8702}
\author{Yu Zhou}
\authornotemark[1]
\orcid{0009-0003-4635-1324}
\email{yzhou243@connect.hkust-gz.edu.cn}
\affiliation{%
  \institution{The Hong Kong University of Science and Technology (Guangzhou)}
  \city{Guangzhou}
  \state{Guangdong}
  \country{China}}

\author{Qiongyan Chen}
\orcid{0009-0004-0893-7014}
\affiliation{%
  \institution{The Hong Kong University of Science and Technology (Guangzhou)}
 \city{Guangzhou}
  \state{Guangdong}
  \country{China}}
\email{qchen580@connect.hkust-gz.edu.cn}

\author{David Yip}
\authornote{Corresponding Author}
\orcid{0000-0002-1745-4741}
\affiliation{%
  \institution{The Hong Kong University of Science and Technology (Guangzhou)}
  \city{Guangzhou}
  \state{Guangdong}
  \country{China}}
\email{daveyip@hkust-gz.edu.cn}

\renewcommand{\shortauthors}{Lei et al.}

\begin{abstract}
We present our approach to using AI-generated content (AIGC) and multiple media to develop an immersive, game-based, interactive story experience. The narrative of the story, \textit{"Memory Remedy"}, unfolds through flashbacks, allowing the audience to gradually uncover the story and the complex relationship between the robot protagonist and the older adults. This exploration explores important themes such as the journey of life, the profound influence of memories, and the concept of post-human emotional care. By engaging with this AIGC-based interactive story, audiences are encouraged to reflect on the potential role of robotic companionship in the lives of older adults in the future, and to encourage deeper reflection on the complex relationship between artificial intelligence and humanity.
\end{abstract}


\begin{teaserfigure}
  \includegraphics[width=\textwidth]{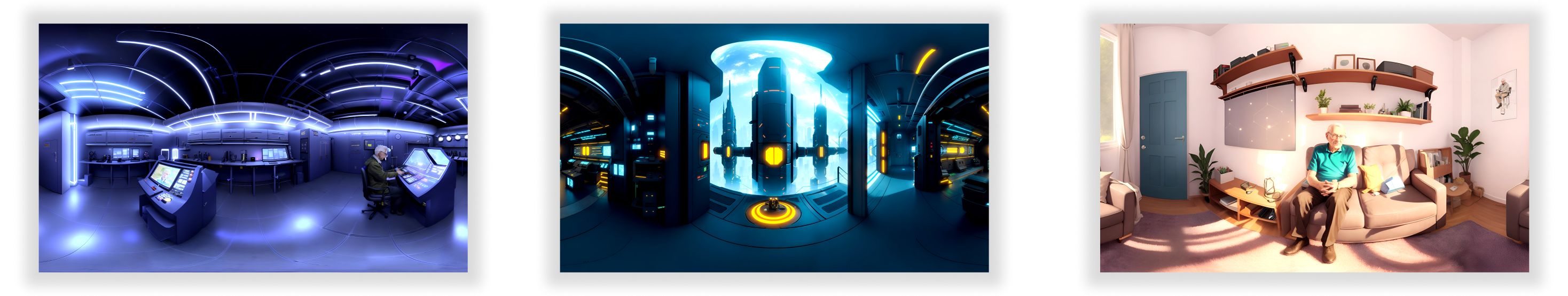}
  \caption{Some examples of interfaces within our interactive story, \textit{Memory Remedy.}}
  \Description{Some examples of interfaces within our interactive story, \textit{Memory Remedy.}}
  \label{fig:teaser}
\end{teaserfigure}


\maketitle

\section{INTRODUCTION}
\textit{Memory Remedy} is an interactive game-based interactive storytelling that explores the themes of aging, connection, and companionship through the lens of artificial intelligence robot. This captivating narrative takes the audiences on a journey of self-discovery, emotional exploration, and profound human connections in a world where AI facilitates memories and companionship.

Drawing from the context of global aging and the devastating impact of Alzheimer's disease (AD). \textit{Memory Remedy} delves into a thought-provoking exploration of how this debilitating condition affects personal and social relationships. The storytelling experience weaves together inter-temporal themes and explores the intricate relationship between humans and AI robot. Through active participation in the first-person perspective, the audience engages with the protagonist's experiences and unusual choices. Through this immersive perspective, viewers gain a deeper understanding of the challenges, emotions, and difficult decisions faced by older adults with Alzheimer's disease. \textit{Memory Remedy} encourages reflection on the profound impact of the disease on personal identity, memory, and the fundamental connections that shape our lives.

AI has proven effective in enhancing interactive storytelling and visual scene generation \cite{ai2024dream360, wang2023simonstown}. We were inspired to investigate if incorporating an innovative AI tool \cite{Skybox}, capable of generating immersive 3D panoramic visual scenes from text, alongside 3D computer graphics using Unreal Engine (UE) \cite{unrealengine}, could enhance the expression of our work.

We continuously refine the hypertext descriptions to craft richer visual scenes and construct a wider variety of narrative paths, to offer audiences a more immersive and dynamic experience. Combining immersive panoramic scenes with a diverse narrative, we intricately designed each choice to facilitate the transition from passive viewing to active decision-making. We found the integration of the AI tool which is effective to create a game-based interactive story that allows for personalized exploration and engagement. 
We provide an overview of the background and motivation behind the creation of \textit{``Memory Remedy''}, discussing the artistic prototype. We proceed to describe the design of the narrative story and the generative pipeline, emphasizing the incorporation of AI technologies at each stage of the production process.
Moreover, we delve into the intricate aspects of hypertext novel story and AI-enhanced scene generation, offering valuable insights into the creative choices and utilization of AI tools throughout the production process. 
We believe that the important contributions of humanities and art and design researchers must be better integrated into human interaction and media discussions of social robots and geriatric care.

\section{BACKGROUND AND MOTIVATION}

\subsection{Story Background}
The most recent data indicate that, by 2050, the prevalence of dementia will double in Europe and triple worldwide, and that estimate is 3 times higher when based on a biological (rather than clinical) definition of Alzheimer's disease\cite{scheltens2021alzheimer}. As economic and social pressures increase with aging populations, countries worldwide are exploring the potential of social, assistive and commitment robots as caregivers and companions for older adults. A growing body of research reflects this interest.

There are many popular movies that explore the theme of companionship between humans and robots. For example, the movie \textit{Robot Frank} (2012) centers around the friendship and companionship between a retired old adult and his companion robot that has given him a new meaning of life in both good and bad way.
The \textit{Robot Revolution} tells the story of a programmer who is dedicated to developing powerful artificial intelligence technology in the hope of enabling human space colonization and exploration, but his efforts also spark concerns about the potential dangers of AI.
The sci-fi film \textit{Ex Machina} (2015) revolves around the emotional entanglement between a young programmer and an intelligent robot. The story explores the role of robot companionship in human emotions and social needs, sparking profound reflections on artificial intelligence and human relationships.
The film \textit{Her} (2013) tells a love story between a man and his intelligent operating system. Through robot companionship, the movie delves into themes of loneliness, human relationships, and emotional dependence.

\subsection{Artistic Style}
Interactive storytelling also encourages exploration and discovery. audiences can uncover hidden storylines, clues, and concepts by solving puzzles, exploring environments, and interacting with characters. This active involvement encourages audiences to think critically and observe details, leading to a deeper understanding of the story and concepts being conveyed.
In addition, interactive storytelling provides real-time feedback and opportunities for audience participation. audiences' decisions and actions directly affect the development of the story and relationships within the game, resulting in immediate feedback and consequences. This sense of agency and impact increases audience engagement and satisfaction. audiences can actively participate in the story, interact with the concepts presented, and gain a better understanding and experiential grasp of them.

\section{RELATED WORK}
As advancements continue in computer graphics and AI technologies, there has been increased emphasis on leveraging AI to dynamically generate visual scenes and seamlessly integrate this content to enhance interactive storytelling experiences \cite{10.1145/3643834.3661547, 10.1145/1236224.1236232}. Most interactive storytelling game systems developed to date have relied predominantly on AI techniques that focus on either plot generation or real-time character behavior control \cite{simonov2019applying, sun2023language}.

 
\subsection{AI in Visual Scene Generation}
Visual scene generation aims to create coherent and contextually relevant images based on textual or visual inputs. This field has seen significant advances with the integration of deep learning techniques, particularly Generative Adversarial Networks (GANs) and Variational Autoencoders (VAEs) \cite{goodfellow2020generative, kingma2013auto}. Large-scale models like DALL-E and Imagen marked a milestone in scene generation by using Transformer architectures to generate detailed images from textual prompts \cite{ramesh2021zero, saharia2022photorealistic}. These models leverage vast datasets and sophisticated architectures to achieve unprecedented realism and detail. Beyond 2D images, generating 3D scenes is another significant frontier. Neural Radiance Fields have successfully synthesized novel views of 3D scenes from sparse 2D images \cite{song2023expanded, mildenhall10representing}. Additionally, the use of panoramas has been proven effective in creating high-quality 3D visual scenes from text input \cite{ai2024dream360, Skybox}.  

\subsection{AI-Facilitated Interactive Storytelling}
AI-powered interactive storytelling is a popular research area where AI is used to create dynamic story experiences that can adapt to user input. Diverse projects demonstrate the trans-formative potential of AI in enhancing Interactive Storytelling \cite{mateas2005structuring, evans2013versu, wang2023simonstown}. 
For instance, \textit{Façade} exemplifies the intersection of AI and interactive drama by combining natural language processing, drama management, and autonomous agents to create a compelling interactive experience \cite{mateas2005structuring}. Additionally, Versu, an interactive storytelling platform, employed social modeling to generate emergent storylines based on audience interactions \cite{evans2013versu}. 
Furthermore, the game \textit{1001 Nights V2} explores the potential of generative AI to enable more meaningful gameplay experiences. It combines the capabilities of instructive language models with advanced image generation techniques. The game organically unites the processes of story crafting and interactive gameplay \cite{sun2023language}.

\subsection{AI-Robot Companionship with Older Adults}
Previous research on AI-robot companionship with older adults has primarily focused on practical robotic interactions and the potential benefits of these technologies. For instance, Abdi et al. conducted a scoping review on the use of socially assistive robot technology in elderly care, emphasizing the potential for these technologies to support daily living activities and provide companionship \cite{abdi2018scoping}. Šabanović et al. took a participatory design approach, involving older adults with depression in the design process of socially assistive robots \cite{vsabanovic2015robot}. However, there is limited research on using AI to design an immersive, game-based, interactive story experience to explore the topic of AI-robot companionship with older adults.

\section{STORY DESIGN AND GENERATIVE PIPELINE}
Our story aims to explore the themes of aging and companion robots in a future society. The narrative focuses on the interplay between memory and choice, presented in a branching structure.
In terms of writing style, we envision incorporating elements of vivid imagination and symbolism. Our primary goal is to explore the relationship between older adults and robots, highlighting the deep connection between personal experience and the shaping of societal destinies. Through this exploration, we hope to elicit empathy from our audience and encourage them to reflect on the importance of individual moral choices and the profound impact they can have.

\section{TRANSFER OF NOVEL STORY}
Our first step was to write a hypertext story as a means of communicating our story concept and creative direction. Hypertext storytelling is embodied in a unique technological environment that allows the audience, as the decision maker of the story, to become more immersed in the development of the story \cite{bell2024reading}. Our work is originally presented in the form of a traditional novel. The novel is then edited and transformed into a storyboard. The storyboard imitates the setting of a movie. The two text forms reflect the characteristics of different media to achieve the transformation between them.

\subsection{Novel Design}
The story is told through flashbacks. The audience is introduced to the story from the first-person perspective of the robot. ``You'' as a robot witness the clue of a delivered package when you open the door, which triggers the entire narrative. You embark on a journey to find your old friend, although you have no knowledge of your relationship with him. 
In the process of guiding the older adult, you revisit the significant places where important moments in their life took place. Seeing AI-generated fragments of the older adult's past memories, such as weddings, funerals, and gardens, makes everything feel so real, yet unverifiable.
As the story progresses, the identity of the old man who sacrificed much in the pursuit of scientific research is revealed. However, it was all worth it, as the invention of the remarkable nostalgic companion robot not only helps the old adult remember many forgotten stories, but also brings hope to society as a whole.
The story highlights the value of memory. Through AI-generated memory fragments, the old man relives significant moments and cherished memories from the past. We do not have a definitive answer as to whether these memories in the story actually happened or were created by ``you'', the robots. 

\begin{figure*}[ht]
  \includegraphics[width=\textwidth]{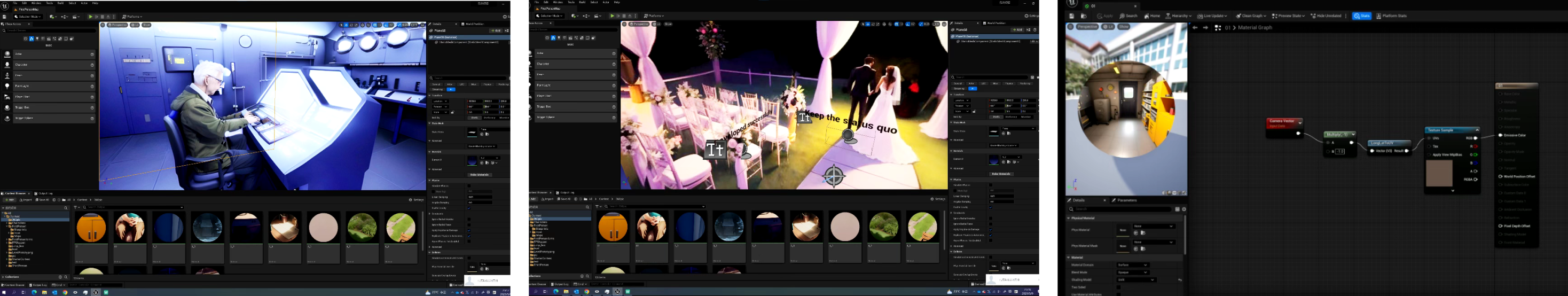}
  \caption{The process involves constructing and refining camera positions in Unreal Engine (UE) to ensure accurate placement and smooth functionality. Additionally, thorough debugging is conducted to identify and address any issues or inconsistencies that may arise during this process.}
  \Description{Building and debugging camera positions in UE.}
  \label{fig:1}
\end{figure*}

\subsection{Storyboard Design}
Interactive storytelling is evolving from the traditional text-based novel to a hypertextual interactive narrative experience that enhances and organizes the storytelling process \cite{yip2022between}. 
We have integrated multiple layers of storylines within the hypertext framework to create multiple endings. This non-linear form of interactive storytelling also coincides with the theme of memory, which is also non-linear in nature. The story is structured in three levels: \textit{Visit, Journey, and Decision,} which create a compelling sense of suspense. The second and third layers of the narrative revolve around the protagonist's memories, delving into his past personal experiences and personal journey. These layers add depth and complexity to the story, providing opportunities for exploration and revelation.

\section{AI-facilitated Scene Generation}
Once the social context, structure, and plot of the story are defined, our attention shifts to the major scenes and scene construction as we seek to translate and adapt the hypertext into visual form. We not only incorporate plot branching to enhance interactivity, but also use AI techniques to create a 360° panoramic world. This allows us to develop immersive and interactive visuals that provide virtual navigation within the interactive story within the scenes. 

One method employed is \textit{Skybox AI} \cite{Skybox}, an AI-driven text-to-panorama tool developed by Blockade Labs for online generation and synthesis of 360° panoramic images (refer to Figure \ref{fig:teaser}). This AI-driven tool can generate panoramic images from regular text input. 
\textit{Skybox AI} not only generates panoramic images, but also allows you to modify and adjust the generated images. It offers two modes: Sketch and Remix. In Sketch mode, we can doodle on the image with a brush and make specific changes to details. In Remix mode, we can change the style and atmosphere of the image. After a few rounds of iteration with \textit{Skybox AI}, we were done with all the scenes in the story.

\section{GAME-BASED INTERACTIVE STORY}
In this section, our primary focus is to seamlessly integrate diverse hypertext and visual elements into a cohesive and impactful game-based interactive story. We referenced relevant game-based storytelling experiences from the past \cite{wei2023interactive}. By utilizing the capabilities of \textit{Unreal Engine} during both production and post-production, we highlight the expressive and communicative potential of the AI-enhanced medium. In addition, we achieve a smooth transition from discrete narrative elements to a fully immersive experience, increasing audience engagement and immersion.

\subsection{Production Details}
In this section, we conducted experiments to explore the immersive and interactive capabilities of AI-generated 2D HDRI (High Dynamic Range Imaging) scenes in the 3D platform of Unreal Engine (UE). For instance, we ingeniously employed camera direction reversal in UE to enhance the three-dimensional perception of the AI panoramic image within the material ball, resulting in a remarkably realistic virtual environment (see Figure.\ref{fig:1}).
Furthermore, to maintain a fixed and static scene, we minimized the number of objects by utilizing planes to construct a cube. This design allows audiences to seamlessly navigate between different scenarios using provided cues and prompts. Interestingly, due to the reversed camera of the material, the panoramic picture observed by audiences from a first-person perspective within the cube appears continuous and realistic. By placing corresponding jump nodes on the plane objects with the material, we achieved smooth scene transitions.
This approach also resulted in an incredibly small project file size (Unreal Engine directly outputs a Windows executable file of only 347MB without compression). Moreover, it requires minimal computing resources while delivering high rendering speeds (achieving above 60 frames per second on an i5-6770 without a GPU).

By integrating the various interactive scenes of the storyline in UE, audiences can dynamically select different levels in real-time by clicking on the available options to immerse themselves in the story. 
Scene transitions are enhanced by synchronized music and voice-over prompts that collectively elevate the overall interactive experience. AI voice-over acting was utilized, with parameters adjusted to align the voice acting with mission characteristics. Dialogue is triggered when audiences encounter characters in different scenes, and selective background music is employed to convey the atmosphere of the environment.
We place great emphasis on contextual and temporal changes between scenes through the use of visual cues and prompts, such as symbolic turnarounds, effect filters, and prominent subtitles accompanied by sound effects. These techniques are designed to increase the viewer's engagement and understanding of the narrative flow.
Through these post production techniques, the audio and visual elements of the real-time interactive story are further enhanced, creating a more immersive and engaging experience for the audience.

\section{DISCUSSION}
After the overall design and deployment of the work were completed, we gathered 13 audiences to participate in the experience of this interactive game, and all of them gave a positive affirmation of the design expression of the work. One audience member mentioned: \textit{``It's a story of memory and choice. I was deeply moved and inspired by it.''} Overall, we found that our work \textit{``Memory Remedy''} enables an immersive, engaging, and memorable storytelling experience through the combination of AI scene-setting and UE. The game-based interactive story presented in this paper explores the potential future directions of artificial intelligence technology scenario building and real-time interactive games for UE, particularly in terms of real-time interaction and storytelling.

We see a future where AI-powered storytelling reaches new levels of immersive, adaptive, and personalized. By analyzing user preferences and behavior, AI algorithms have the potential to dynamically customize stories for individual users, resulting in unique and tailored experiences. In this project, our goal is to merge AI-generated scenes with immersive experiences by harnessing the power of UE, augmented reality (AR), and virtual reality (VR) technologies. This integration enables the audience to gain a deeper understanding of the narrative and to explore the story within an interactive and immersive environment.
This endeavor pushes the boundaries of storytelling, enabling a new level of collaboration between AI scene generation and virtual scene fusion. As a result, interactive storytelling projects are transformed into immersive experiences that captivate and engage audiences on a whole new level.

\section{CONCLUSION}
``\textit{Memory Remedy}'' is an immersive narrative experience that seamlessly blends the themes of aging, connection, and companionship with the advancements of AI technology. Through game-based interactive play and captivating storytelling, its objective is to elicit empathy, introspection, and a profound understanding of the human experience. It offers an unforgettable journey of self-discovery and exploration, where cherished memories, meaningful connections, and the transformative power of companionship are unveiled. Additionally, during the post-production stage, we experimented with and innovated the integration of AI tools into the interactive storytelling experience.



\bibliographystyle{ACM-Reference-Format}
\bibliography{mainreference}
\end{document}